\documentclass[prl,twocolumn,superscriptaddress]{revtex4-2} 
\usepackage[T1]{fontenc}
\usepackage[utf8]{inputenc}
\usepackage{lmodern}                 
\usepackage{microtype}               
\usepackage{mathrsfs,amsfonts, amsmath,amssymb,mathtools}
\usepackage{txfonts}
\usepackage{graphicx,subfigure}
\usepackage{bm}
\usepackage{color}
\usepackage[normalem]{ulem}
\usepackage{dcolumn} 
\usepackage{bbold}
\usepackage[table]{xcolor}
\usepackage{hyperref}
\usepackage{booktabs}
\usepackage{url}
\usepackage{siunitx}                 
\usepackage{xspace}                  
\usepackage{physics}                 

\hypersetup{
  colorlinks=true,
  linkcolor=purple,  % \ref, \eqref, figure/table/section links
  citecolor=purple,  % \cite / biblatex citations
  urlcolor=purple,   % \url and \href URLs
  filecolor=purple
}

\begin{document}
\title{Ferroelectrical Switching as a Probe of Quantum Damping in Magnetic Spin Systems}

\author{Yuefei Liu} \email{yuefei.liu@ntnu.no}
\affiliation{Department of Physics and Astronomy, Uppsala University, Box 516, SE-751 20 Uppsala, Sweden}
\affiliation{Department of Applied Physics, School of Engineering Sciences, KTH Royal Institute of Technology, AlbaNova University Center, SE-10691 Stockholm, Sweden}
\affiliation{Nordita, Stockholm University and KTH Royal Institute of Technology, SE-10691 Stockholm, Sweden}
\affiliation{Center for Quantum Spintronics, Department of Physics, Norwegian University of Science and Technology NTNU, NO-7491 Trondheim, Norway}
\author{Anna Delin}  
\affiliation{Department of Applied Physics, School of Engineering Sciences, KTH Royal Institute of Technology, AlbaNova University Center, SE-10691 Stockholm, Sweden}
\affiliation{Swedish e-Science Research Center (SeRC), KTH Royal Institute of Technology, SE-10044 Stockholm, Sweden}
\affiliation{Wallenberg Initiative Materials Science for Sustainability (WISE), KTH Royal Institute of Technology, SE-10044 Stockholm, Sweden}
\author{Olle Eriksson}  
\affiliation{Department of Physics and Astronomy, Uppsala University, Box 516, SE-751 20 Uppsala, Sweden}
\affiliation{WISE-Wallenberg Initiative Materials Science, Uppsala University, Box 516, SE-751 20 Uppsala, Sweden}
\author{Erik Sj\"oqvist} 
\affiliation{Department of Physics and Astronomy, Uppsala University, Box 516, SE-751 20 Uppsala, Sweden}
\author{Kaiyou Wang}
\affiliation{State Key Laboratory of Superlattices and Microstructures, Institute of
Semiconductors, Chinese Academy of Sciences, Beijing 100083, China}
\affiliation{Center of Materials Science and Optoelectronics Engineering, University of Chinese
Academy of Sciences, Beijing 100049, China}
\author{Qirui Cui} \email{qiruic@kth.se}
\affiliation{Department of Applied Physics, School of Engineering Sciences, KTH Royal Institute of Technology, AlbaNova University Center, SE-10691 Stockholm, Sweden}
\affiliation{Swedish e-Science Research Center (SeRC), KTH Royal Institute of Technology, SE-10044 Stockholm, Sweden}

\date{\today}

\begin{abstract}
While damped spin dynamics is important for the understanding of magnetic materials, clear signatures of \emph{quantum corrections} to the Gilbert damping mechanism remain elusive.
We propose a route to distinguish quantum and classical Gilbert spin damping using ferroelectric control of a magnetic dimer. 
\emph{Ab initio} calculations for dimers on ferroelectric substrates show that polarization reversal switches the inter-spin exchange between ferromagnetic and antiferromagnetic regimes.
We formulate a magnetization-based diagnostic that relates magnetization traces to entanglement dynamics, which enables ferroelectrical on/off control of dimer entanglement.
Material-informed quantum Landau-Lifshitz-Gilbert simulations illustrate how the signature of magnetization dynamics can, in principle, be used to infer the existence of quantum Gilbert spin damping.
This minimal and non-volatile platform connects first-principles modeling to experimentally accessible observables and provides a starting point for voltage-controlled quantum entanglement in magnetic spin networks.
\end{abstract}

\maketitle

Gilbert damping plays a central role in the description of how spins exchange energy and information with their environment. 
Classically, this process is captured by the Landau–Lifshitz–Gilbert (LLG) equation, in which spin damping emerges from spin–orbit and bath couplings \cite{Eriksson2017}.
Beyond being central to understanding spin dynamics in materials, the LLG equation exhibits rich nonlinear dynamics and mathematical structure~\cite{Lakshmanan2011}.
At the quantum level, several strands have emerged, such as links between Lindblad dynamics and spin damping \cite{gaitan24,uhrig25}, as well as extensions such as generalized or fractional LLG equations~\cite{verstraten23}.
More recently, nonlinear master equations have been proposed, which incorporate damping-like behavior while preserving quantum coherence of the spins \cite{Wieser2013,liu24}.
Despite these developments, a direct, experimentally accessible existence proof of the uniquely quantum mechanical aspects of spin dynamics has yet to be achieved. 

Quantum aspects of magnetic spin systems have attracted intense interest across theory and experiment.
On the theoretical side, recent efforts align with the goals of quantum magnonics~\cite{yuan22};
for example, antiferromagnetic magnon-squeezed states have been theoretically predicted to exhibit intrinsic entanglement arising from energy minimization and strong inter-sublattice interactions~\cite{kamra20}.
By coupling a qubit to equilibrium magnon squeezing, one can generate and probe two-mode squeezed states~\cite{romling23,romling24}, and open-system treatments based on the Lindblad master equations quantify relaxation and dephasing of such magnon-based entangled states~\cite{yuan22L,yuan22ME}.
On the experimental side, inelastic neutron scattering now enables entanglement verification by mapping the dynamical structure factor to the quantum Fisher information, certifying multipartite entanglement in frustrated quantum spin liquids at finite temperature \cite{scheie21}.
Ferroelectric (FE) switching in FE–magnetic heterostructures (HS) offers a current-driven and nonvolatile method to control magnetism by tuning spin couplings. 
This control can arise from polarization-induced charge redistribution that modifies interfacial magnetism~\cite{me1,gong1,gong2}. 
Beyond electrical switching, similar control can be actuated mechanically via flexoelectricity~\cite{fe2} and optically by ultrafast laser excitation~\cite{fe3}.
Such a recent effort on electric field control of magnetization has provided a path to pursue electrical control of information processing and storage in magnetic spin systems~\cite{Fert2024}.

By taking advantage of the progress in quantum LLG ($q$-LLG) theory and FE-control of spin couplings, this Letter focuses on Gilbert-type quantum damping of magnetic spin systems.
We investigate such physics by changing the quantum entanglement of magnetic dimers using FE switching in FE-magnetic HS. 
Our theoretical investigation finds that FE control of magnetic dimer couplings provides a practical route to demonstrate the quantum nature of Gilbert spin damping.

\begin{figure}
    \centering
    \includegraphics[width=0.98\linewidth]{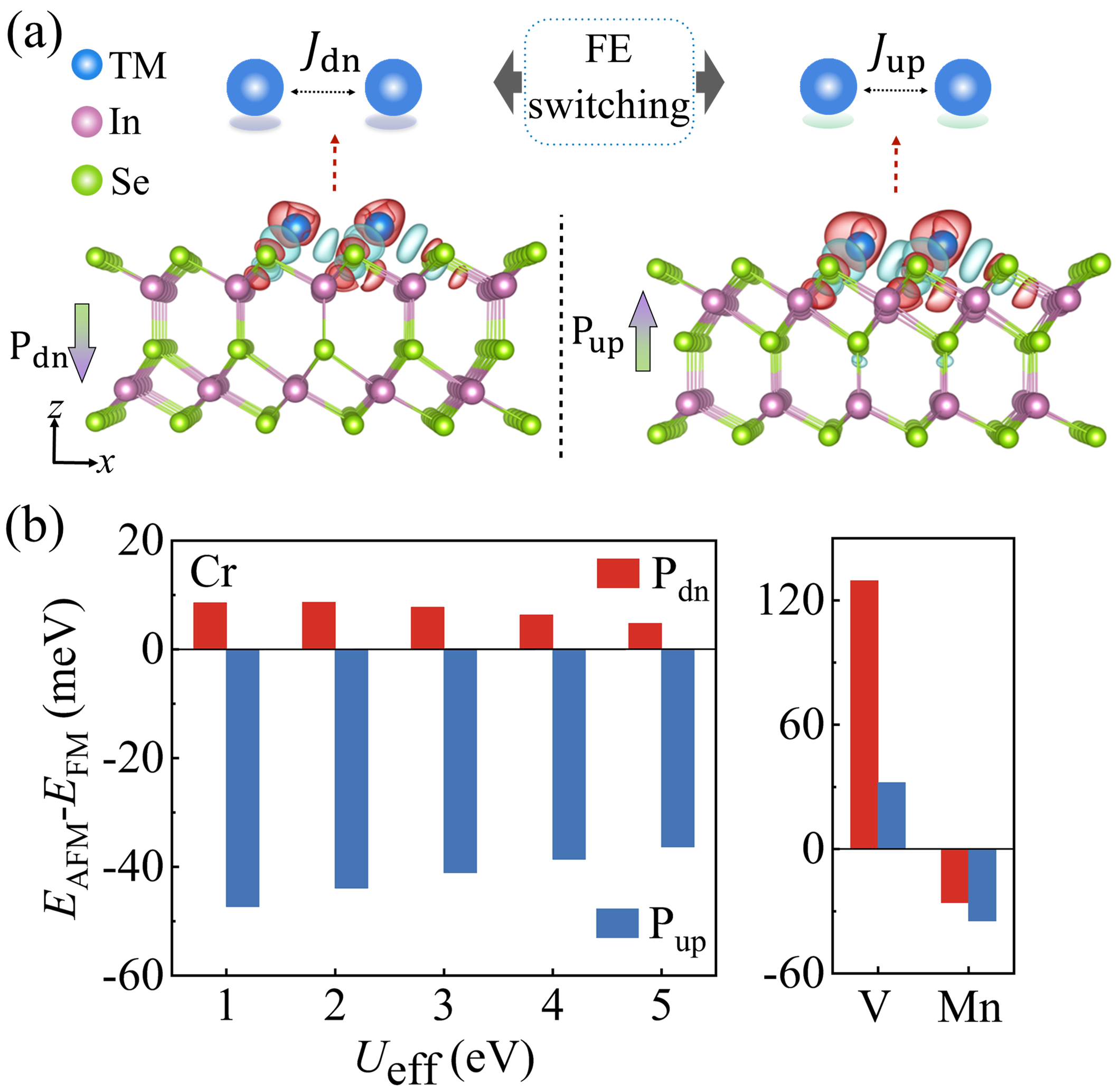}
    \caption{The ferroelectricity tunes the magnetic dimer coupling strength from $J_{\rm up}$ to $J_{\rm dn}$, as the  In$_{2}$Se$_{3}$ substrate polarization switches \textit{from upward to downward}, as illustrated in (a).
    The schematics of the crystal structure of the magnetic transition metal (TM) dimer on the In$_{2}$Se$_{3}$ substrate with \textit{downward} polarization $\rm P_{dn}$ and \textit{upward} polarization $\rm P_{up}$. 
    For Cr, The out-of-plane electric polarization reaches -0.054 eÅ/u.c.\,in the downward state and 0.116 eÅ/u.c.\,in the upward state.
    The bonding between Cr and In$_2$Se$_3$ can be seen from the charge-density difference,
    $\Delta \rho = \rho(\mathrm{HS}) - \rho(\mathrm{Cr}) - \rho(\mathrm{In}_2\mathrm{Se}_3)$,
    as shown by the contour maps at the isovalue
    $\Delta \rho = \pm 2 \times 10^{-3}\,\mathrm{e}/\AA^{3}$.
    (b) For the Cr dimer, the ferroelectricity-controllable energy difference between ferromagnetic (FM) and antiferromagnetic (AFM) phases as a function of Hubbard $U_{\mathrm{eff}}$ correction. With $U_{\mathrm{eff}}= 2$ eV, we show the ferroelectricity-controllable energy difference between FM and AFM phases for the V dimer and Mn dimer.}
    \label{fig:P-up-dn}
\end{figure}

We model the magnetic dimer by the spin Hamiltonian
\begin{eqnarray}
\begin{aligned}
H = &\sum_{\beta=x,y,z} J_{12}^{\beta\beta} S_1^\beta S_2^\beta +\sum_{j=1, 2}\left[K_1(S_j^x)^2+K_2(S_j^y)^2\right]
 \\
+ & \boldsymbol{D} \cdot (\boldsymbol{S}_1 \times \boldsymbol{S}_2)-\sum_{j=1,2} \gamma_g \boldsymbol{B} \cdot \boldsymbol{S}_j,
\end{aligned}
\label{eq:H}
\end{eqnarray}
where we use the same symbol $\mathbf{S}_j$ in the classical and quantum formulations, and the gyromagnetic ratio $\gamma_g = g\mu_B / \hbar$ with dimensionless g-factor and $\mu_B$ Bohr magneton.
Hereafter, we set $\hbar = 1$. 
Classically, $\mathbf{S}_j\in\mathbb{R}^3$ has fixed magnitude $\mu_j$ and we write $\mathbf{S}_j=\mu_j \mathbf{m}_j$, where $\mathbf{m}_j$ is a unit vector that describes the direction of the classical spin at site $j$.
Quantum mechanically, $\mathbf{S}_j=(S_j^x,S_j^y,S_j^z)$ denotes the vector of spin operators. 
Its components can be viewed as $(2s+1)\times(2s + 1)$ Hermitian matrices. The Heisenberg exchange couplings $J^{\beta \beta}$, $\beta = x, y, {\rm or}, z$, single-ion magnetic anisotropy $K_1 (K_2)$ on site $1 (2)$ in the dimer, Dzyaloshinskii-Moriya (DM) interaction $\mathbf{D}$, and the external magnetic field $\mathbf{B}$ are explicitly included in 
the spin dynamics simulation.
The magnetic transition metal (TM) dimers have been considered in the FE tunable system, which is illustrated in Fig.~\ref{fig:P-up-dn}~(a) and calculated within the framework of density functional theory (DFT), as implemented in VASP \cite{sm2, sm3, sm4}. 

The exchange-correlation energy is treated using the generalized gradient approximation in the Perdew–Burke–Ernzerhof functional form \cite{sm5}. To properly account for strong on-site Coulomb interaction associated with localized 3$d$ electrons of Cr, the Hubbard $U_{\rm eff}$ correction is employed. A plane-wave energy cut-off is set to 400~eV, and the Brillouin zone for the two-dimensional structure is sampled using a $\Gamma$-centered $3\times3\times1$ $k$-point mesh. A vacuum layer of more than 15~Å is included along the out-of-plane direction to eliminate interaction between periodic layers. The structures are optimized until the Hellmann–Feynman forces are less than 0.01 eV/Å. 

Figure~\ref{fig:P-up-dn}~(a) schematically illustrates the modulation of exchange coupling in a magnetic TM dimer via FE switching. The dimer resides on the surface of an FE substrate, where FE switching reconfigures the surface electrostatic potential and charge density. 
Consequently, the orbital hybridization between the dimer and the surface ligands is modified.
This variation in wave-function overlap can alter the magnitude, and potentially the sign, of the ligand-mediated superexchange within the dimer.
Figure~\ref{fig:P-up-dn}~(a) also shows the side-view crystal structures of a magnetic TM dimer on the ferroelectric In$_{2}$Se$_{3}$ substrate \cite{sm6, sm7, sm8}.
Among the magnetic transition-metal adatoms considered, including V, Cr, and Mn, the Cr dimer shows the clearest FE-controlled sign switching of the exchange coupling in our calculations.
In the most stable configuration, the Cr atoms are located at hollow sites of the surface (see End Matter).
To ensure isolated dimer behavior, we use a 5$\times$5$\times$1 In$_{2}$Se$_{3}$ supercell to minimize interaction between adjacent dimers (see End Matter)~\cite{a1}.
The magnetic moment of each Cr atom is approximately 4.3 $\mu_{B}$, where the closest localized spin configuration corresponds to $s=2$.
Reversing In$_2$Se$_3$ from the down ($\text{P}_{\mathrm{dn}}$) to the up ($\text{P}_{\mathrm{up}}$) polarization, switches the magnetic ground state of the Cr dimer. Specifically, at Hubbard $U_{\mathrm{eff}}=2$ eV, $\text{P}_{\mathrm{dn}}$ stabilizes the ferromagnetic (FM) phase with $\Delta E \equiv E_{\mathrm{AFM}} - E_{\mathrm{FM}} = 8.66$~meV, whereas $\text{P}_{\mathrm{up}}$ yields the antiferromagnetic  
(AFM) phase with $\Delta E = -43.93$~meV [Fig.~\ref{fig:P-up-dn}~(b)]. $E_{\mathrm{AFM}(\mathrm{FM})}$ represents the self-consistent energy of the Cr dimer in AFM (FM) coupling configuration.
This FM-AFM transition is robust over a wide range of $U_{\rm eff}$ values (1–5 eV).
Under identical structural configurations, we further replace Cr with V or Mn and examine the effect of polarization reversal ($\text{P}_{\mathrm{dn}}\!\to\!\text{P}_{\mathrm{up}}$) on $\Delta E$. $\Delta E$ changes from $129.50$ to $32.05$~meV for the V dimer and from $-25.84$ to $-34.16$~meV for the Mn dimer [Fig.~\ref{fig:P-up-dn}~(b)]. 
These variations indicate a strengthened AFM exchange under $\text{P}_{\mathrm{up}}$, while the magnetic ground state remains unchanged.

\begin{table}[htbp!]
\centering
\setlength{\tabcolsep}{2.8pt} 
\renewcommand{\arraystretch}{1.2}
\begin{tabular}{ccccccccc}
\toprule
 & \raisebox{0.3ex}{$J_{xx}$} & \raisebox{0.3ex}{$J_{yy}$} & \raisebox{0.3ex}{$J_{zz}$} & \raisebox{0.3ex}{$K_1$} & \raisebox{0.3ex}{$K_2$} & \raisebox{0.3ex}{$D_x$} & \raisebox{0.3ex}{$D_y$} & \raisebox{0.3ex}{$D_z$} \\
\hline
\raisebox{-0.5ex}{P$_{\text{dn}}$} & \raisebox{-0.5ex}{-1.096} & \raisebox{-0.5ex}{-1.077} & \raisebox{-0.5ex}{-1.080} & \raisebox{-0.5ex}{-0.096} & \raisebox{-0.5ex}{-0.094} & \raisebox{-0.5ex}{-0.005} & \raisebox{-0.5ex}{-0.095} & \raisebox{-0.5ex}{0.169} \\[5pt]
\raisebox{0.1ex}{P$_{\text{up}}$} & \raisebox{0.1ex}{5.497} & \raisebox{0.1ex}{5.495} & \raisebox{0.1ex}{5.492} & \raisebox{0.1ex}{0.004} & \raisebox{0.1ex}{-0.002} & \raisebox{0.1ex}{-0.002} & \raisebox{0.1ex}{-0.075} & \raisebox{0.1ex}{0.020} \\[0pt]
\bottomrule
\end{tabular}
\caption{DFT‑calculated magnetic parameters, including exchange couplings $J$, single-ion magnetic anisotropy $K$ and DM interaction $D$, under opposite ferroelectric polarizations. All values are in meV.} 
\label{tab:parameters-I}
\end{table}

\begin{figure}% [h!]
    \centering
    \includegraphics[width=0.95\linewidth]{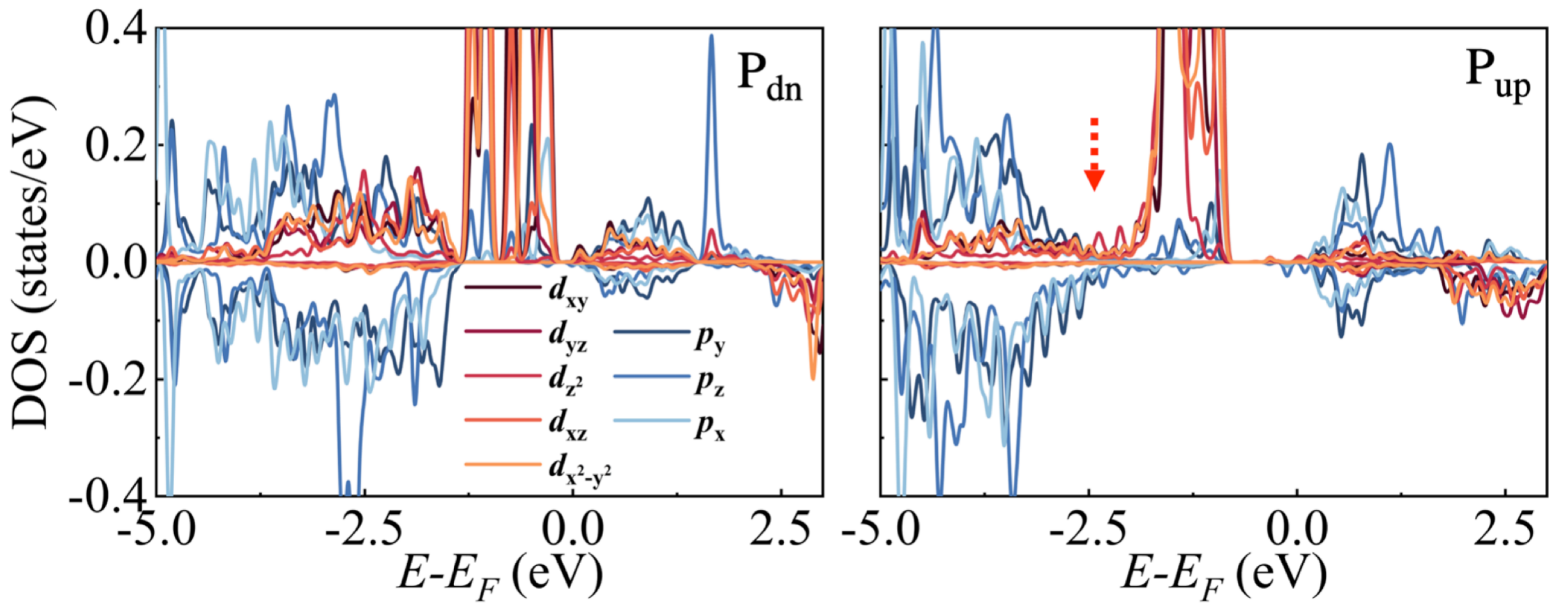}
    \caption{The projected density of states (DOS) of Cr dimer on In$_2$Se$_3$.
    The red dotted arrow is to emphasis that the $p-d$ orbital hybridization is weakened when ${\rm P}_{\rm up}$ up polarization.
    }
    \label{fig:FE_DOS}
\end{figure}

According to the Goodenough--Kanamori--Anderson (GKA) rules \cite{sm9,sm10,sm11}, the interfacial Cr--Se--Cr angle of $\sim99^\circ$ lies near the $90^\circ$ limit and thus tends to favor FM superexchange, while direct Cr--Cr exchange favors AFM alignment, leading to competition \cite{ssm11}. 
Reversing the out-of-plane polarization flips the sign of the bound charge at the top surface where the Cr dimer resides: 
for $\text{P}_{\text{dn}}$ the surface carries negative bound charge, whereas for $\text{P}_{\text{up}}$ it is positive. 
This sign reversal modulates the interfacial carrier density and shifts the local electrostatic potential, thereby tuning the Cr–Se hybridization strength. 
Specifically, $\text{P}_{\text{dn}}$ produces higher electron density around the interfacial Cr/Se and thereby strengthens the $p$--$d$ orbitals hybridization. 
The charge accumulation is evident in the charge-difference isosurfaces [Fig.~\ref{fig:P-up-dn}~(a)], and the enhanced hybridization is corroborated by the projected density of states (DOS), as shown in Fig.~\ref{fig:FE_DOS}, in End Matter. 
In contrast, $\text{P}_{\text{up}}$ depletes electrons from the bonding region, rendering the bond more ionic, and thus weakens the hybridization.
Consequently, FM superexchange is stronger for $\text{P}_{\text{dn}}$, whereas under $\text{P}_{\text{up}}$ the reduced hybridization allows AFM exchange to prevail, accounting for the ferroelectric polarization-controlled transition between FM and AFM ground states shown in Fig.~\ref{fig:P-up-dn}~(b).

In general, our ferroelectrically tuned coupling scheme $\rm P_{up} \leftrightarrow P_{dn}$ admits three typical modes: (i) intra-FM tuning (varying the exchange within the FM regime), (ii) intra-AFM tuning (varying it within the AFM regime), and (iii) FM$\leftrightarrow$AFM switching via a sign change of the effective exchange $J$.
Employing energy-mapping approaches \cite{ssm12, sssm12}, we determine the spin Hamiltonian parameters in Eq.~\eqref{eq:H} for Cr dimers substrate on In$_{2}$Se$_{3}$  in Table~\ref{tab:parameters-I}. The exchange couplings are found to be approximately isotropic \cite{sm12}.
Notably, switching the polarization direction of In$_2$Se$_3$ from downward to upward distinctly tunes the  FM exchange coupling from $\sim\!-1.1$~meV to an AFM coupling of $\sim\!5.5$~meV, 
which provides a promising route for electrically controlled on–off switching of entanglement as demonstrated by the following q-LLG simulations.

\begin{figure*}[ht!]
    \centering
    \includegraphics[width=0.92\linewidth]{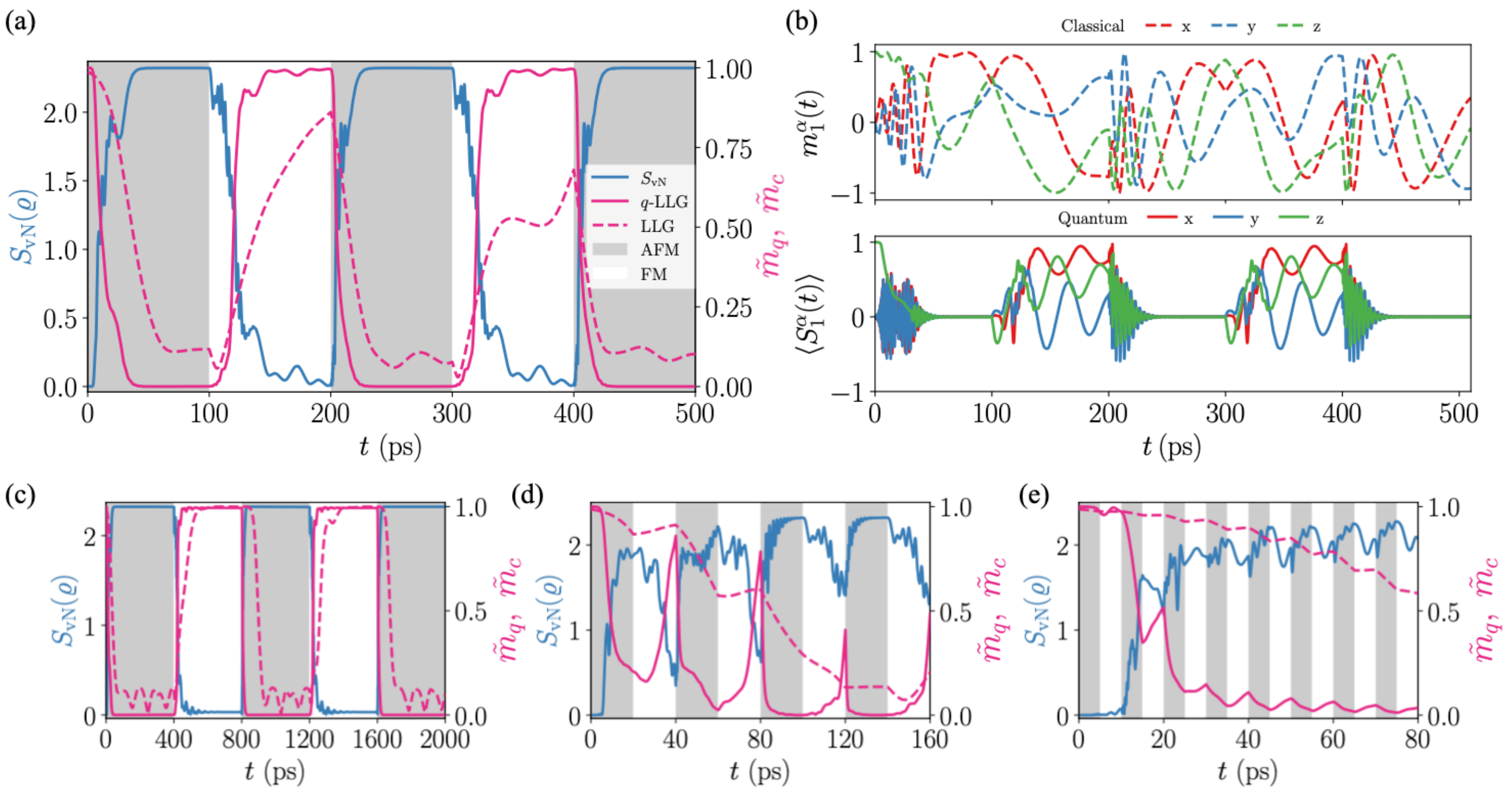}
    \caption{(a) $q$-LLG time evolution of the von Neumann entropy $S_{\mathrm{vN}}(\varrho)$ as well as the $q$-LLG ($\tilde{m}_q(t)$) and LLG ($\tilde{m}_c(t)$) average magnetic moment per Cr atom, for a Cr dimer on an $\mathrm{In_2Se_3}$ substrate under FE switching. Simulations start from the FM product quantum state $\ket{\uparrow\uparrow}$ and corresponding FM classical configuration $\uparrow\uparrow$. At $t=$ 100, 200, 300, 400 ps, the Cr–Cr couplings are switched between the polarization-induced parameter sets $P_{\mathrm{up}}$ (AFM) and $P_{\mathrm{dn}}$ (FM), as listed in Table~\ref{tab:parameters-I}, in the sequence AFM–FM–AFM–FM–AFM. (b) Component-resolved dynamics for one spin in the dimer (the other is mirrored): the $q$-LLG expectation values $\langle S_1^\beta(t)\rangle$ and the LLG magnetic-moment components $m_1^\beta(t)$ for $\beta\in\{x,y,z\}$.
    Under the same simulation settings as in (a), (c)-(e) show the results with switching intervals of 400, 20 and 5 ps. Note, the upper bound of von Neuman entropy in spin-$2$ dimer system is 2.32~\cite{remark-vNEntr}. }
    \label{fig:transitions-protocol}
\end{figure*}

The classical atomistic spin dynamics of the Cr-dimer system can be simulated by LLG equations, which is written as
\begin{eqnarray}
\dot{\mathbf{m}}_j & = & 
-\gamma_g {\bf m}_j \times\mathbf{H}_j
+ \frac{\alpha}{|\mathbf{m}_j|}\,\mathbf{m}_j \times\dot{\mathbf{m}_j},
    \label{eq:LLG}
\end{eqnarray}
where $\alpha$ is the Gilbert damping parameter and $\mathbf{H}_j = - \partial H / \partial {\bf m}_j$, with $H$ the Hamiltonian in Eq.~\eqref{eq:H}, is the local effective field incorporating the net influence of all spin–spin interactions and external drives felt by the simulated spin at site $j$.
The evolution of quantum correlations in a damped magnetic dimer lies beyond the classical LLG framework, but is captured by the $q$-LLG equation.
The $q$-LLG equation takes the form \cite{liu24}
\begin{eqnarray}
\dot{\varrho} = i [\varrho,H] + i \kappa [\varrho,\dot{\varrho}],
\label{eq:qllg}
\end{eqnarray}
with the quantum Gilbert damping parameter $\kappa$ and density operator $\varrho$ of the dimer system.
In our study, the coupling strength is in the 
order of 1 meV, therefore, the time scale of spin dynamics simulations is $\hbar/J \sim 3$ ps.
We take the Gilbert damping to be $\alpha = \kappa = 0.1$ in both LLG and $q$-LLG throughout this work.
The quantum state space of the two spin-2 Cr atoms are spanned by the product states $\ket{2;m} \otimes \ket{2;m'}$, $m,m'=-2,-1,\ldots, 1, 2$, with $\ket{2;m}$ and $\ket{2;m'}$ being the $S^z$ eigenvectors. We consider pure state ($\varrho = |\psi\rangle\langle\psi|$) evolution, for which the von Neumann entropy $S_{\rm vN}$ of the reduced state of one of the Cr atoms is a well-defined entanglement measure \cite{remark-vNEntr}.

Magnetization provides the most direct bridge between spin dynamics and possible experimental readout.
While classical LLG constrains each local magnetic moment to have a fixed magnitude, $q$-LLG evolves the quantum state itself, allowing entanglement and other non-classical correlations to modify the spin expectation values~\cite{liu24}.
In classical LLG dynamics, the time evolution of the average magnetic moment per Cr atom is obtained from the spin-direction vectors ${\bf m}_j$, i.e., $\tilde{m}_c = \frac{1}{2}\left|\sum_{j=1}^{2} {\bf m}_j\right|$. 
Analogously, in the quantum formulation,  we use $\tilde{m}_q = \frac{1}{2}|\sum_j^2 \langle {\bf S}_j \rangle|$ with $\langle S_j^\beta(t) \rangle = \operatorname{Tr} \left[\varrho(t) S_j^\beta\right]$, $\beta=x,y,z$, to quantify the average magnetic moment per Cr atom.
While magnetization probes the local spin expectation values, the dynamics of a quantum dimer state is not determined by these one-spin observables alone.
Entanglement and other nonclassical correlations provide the missing many-body information and can strongly reshape the magnetization dynamics.
Thus, the numerically resulting difference in magnetization dynamics provides a possible signature of quantum corrections to the Gilbert damping mechanism.

Figure~\ref{fig:transitions-protocol} (a) shows the LLG and $q$-LLG time evolution of the average magnetic moment per atom, $\tilde{m}_c$ and $\tilde{m}_q$, respectively, for the Cr dimer using the DFT-derived parameters listed in Table.~\ref{tab:parameters-I}, as well as the entanglement entropy $S_{\mathrm{vN}}$.
We initialize the classical dimer in the FM state  $\uparrow \uparrow$, and the quantum dimer in the state $\ket{2;2} \otimes \ket{2;2}$. 
We apply a uniform magnetic field $\mathbf{B}=B (1,0,1)/\sqrt{2}$, and choose its magnitude $B$ such that the Zeeman energy is comparable to the exchange energy, i.e., $g\mu_B B \sim J$. For $J=1\,\mathrm{meV}$ and $g\approx2$, this corresponds to $B\approx 8.6\,\mathrm{T}$.
The coupling profile in Fig.~\ref{fig:transitions-protocol} (a) is instantaneously switched every 100\,ps according to the sequence AFM–FM–AFM–FM–AFM during the 500-ps simulation.
Note that the exchange rearrangement is treated as instantaneous in our simulations~\cite{nc1,prls1}. This approximation is justified because the FE polarization reversal in In$_2$Se$_3$, driven by collective ionic displacements, is expected to occur on a $\sim$1 ps timescale \cite{nl1}, which is sufficiently fast compared with the spin dynamics considered in our model.
In the classical case, the dimer tends to evolve into either AFM or FM spin alignment, depending on the AFM or FM-type of the exchange coupling.
The observed oscillations in the dynamics arise from the torque switching of the classical angular momentum, and the precession motion that follows from Eq.~\eqref{eq:LLG}. 
In the quantum case, our results show that the level of entanglement is 
inversely linked to the net magnetization of the quantum dimer [Fig.~\ref{fig:transitions-protocol}].  
Such an inverse relationship is most pronounced during switching events, where drops in net magnetization coincide with rapid rises in entanglement, and vice versa.
Here, the magnetization should be viewed as an indirect signature rather than a standalone entanglement measure.
In our simulations, its reduction correlates with the growth of dimer entanglement, but the full quantum state for these $s=2$ atoms also contains correlation information that is not captured by local spin component expectation values alone.
Furthermore, the AFM intervals appear as long, stable plateaus in both magnetization and entanglement, while the FM intervals show rapid fluctuations in both quantities.
These differences trace back to the two coupling profiles themselves. 
The AFM profile is dominated by a strong exchange interaction term, whereas the FM profile features a relatively large DM interaction (e.g., $D_z/J_{zz} \approx 0.16$).
The large DMI in the FM interval is important because it breaks the collinear spin symmetry and introduces anisotropic, non-collinear torques, which can enhance the sensitivity of the magnetization dynamics to quantum correlations.
In this sense, the DFT-calculated DMI in the FE-tunable Cr dimer provides a promising setting for exploring quantum signatures of damped spin dynamics.
Figure~\ref{fig:transitions-protocol}~(b) presents the component-resolved spin dynamics, comparing the classical (LLG) and quantum ($q$-LLG) results, which further illustrates the difference between the LLG and $q$-LLG dynamics.

To further examine the dynamical origin of the LLG/$q$-LLG discrepancy, we compare the two approaches over shorter switching time intervals.
As shown in Fig.~\ref{fig:transitions-protocol} (d) and (e), reducing the FE-switching to $20\,\mathrm{ps}$ and $5\,\mathrm{ps}$ enhances the difference between the classical and quantum magnetization dynamics.
Conversely, the slower-switching case with a $400\,\mathrm{ps}$ interval, shown in Fig.~\ref{fig:transitions-protocol} (c), exhibits a reduced discrepancy.
This trend suggests that faster coupling modulation makes the dynamics more sensitive to quantum correlations, thereby amplifying the signatures captured by the $q$-LLG equation but absent in classical LLG.
We emphasize that this comparison is qualitative: varying the pulse length does not by itself constitute an entanglement signature, but it provides a useful way to test how the quantum character of damping enters the magnetization response.
A natural and potential experimental extension would be to compare engineered spin chains of different lengths, where longer chains are expected to approach more classical LLG-like behavior while shorter chains retain stronger quantum signatures.

To conclude, our calculations identify FE-controlled magnetic TM dimers as a minimal setting in which quantum corrections to the Gilbert damping mechanism can be tested through magnetization dynamics.
The key point is not that magnetization is an entanglement witness by itself, but that, under the proposed switching protocol, its time dependence carries a clear numerical signature of the entanglement-dependent dynamics captured by $q$-LLG but absent in classical LLG.
The sizable DMI found in the FM interval further strengthens this distinction by introducing anisotropic torques and breaking collinear spin symmetry.

More broadly, switchable exchange in an atomically small magnetic unit offers a controlled setting for studying damped quantum spin dynamics, beyond static Hamiltonian engineering.
The dimer is the minimal element of a spin network, so extending this idea to few-spin clusters, finite chains, or coupled dimer arrays would allow one to examine how quantum signatures evolve toward classical behavior with increasing system size.
This perspective also connects naturally to recent interest in memory effects and non-Markovian magnetisation dynamics~\cite{Hartmann2025NonMarkov,Felipe2025}.
Beyond FE polarization reversal, related coupling-control schemes may be pursued through strain gradients, flexoelectricity~\cite{fe2}, or ultrafast optical driving~\cite{fe3}.
On the experimental side, atomically assembled dimers on FE substrates, prepared by STM-based techniques \cite{stm1,stm2,stm3,stm4} and probed by time-resolved magneto-optical methods \cite{ke1,ke2}, would provide a possible route to test the predicted magnetization signatures.
We hope that this work motivates further studies of controllable entanglement, nonlinear damping, and memory effects in quantum spintronic systems.

\vspace{1cm}
\textit{\textbf{Acknowledgements}---}
Y.L acknowledges financial support from Nordita, Grant No. ERC-SYG 81451, the Research Council of Norway (RCN), Grant No. 354571, and the Center of Excellence in Quantum Spintronics, NTNU Norway, RCN Grant No. 262633.
Q. C. and A. D. acknowledge financial support from the Swedish Research Council (VR), Grant Nos. 2019-05304,  2022-04720, and 2024-04986, as well as the Knut and Alice Wallenberg foundation (KAW), Grant Nos. 2018.0060, 2021.0246, and 2022.0108.
A. D. acknowledges support from the Wallenberg Initiative Materials Science (WISE), funded by the Knut and Alice Wallenberg Foundation and the Swedish e-Science Research Centre (SeRC).
O.E. acknowledges financial support from the Swedish Research Council (VR) and the Knut and Alice Wallenberg foundation (KAW).
O.E. acknowledges support from the Wallenberg Initiative Materials Science (WISE), funded by the Knut and Alice Wallenberg Foundation, for support, as well as support from STandUPP, the ERC (FASTCORR project) and eSSENCE.
E. S. acknowledges financial support from the Swedish Research Council (VR), Grant No. 2025-05249.
K. W. acknowledges support from the National Natural Science Foundation of China, Grant Nos. 12427805 and 12241405. The computations are enabled by resources provided by the National Academic Infrastructure for Supercomputing in Sweden (NAISS), partially funded by the Swedish Research Council (VR), under Grant No. 2022-06725.

%Y. L. and Q.C. initiated the discussion and developed the theory framework with A. D., O. E., E. S.and K.W.. 
%Y. L. and Q.C. implemented the calculations of the physical model, and wrote the first draft of the manuscript, and A. D., O. E., E. S.and K.W. supervised the project.
%All authors contributed to the discussions of the results and finalization of the manuscript.

%\textit{\textbf{Data availability}---}
%The entire data set that support the findings of this article is not publicly available.
%The data are available from the authors upon reasonable request.

\newpage

\ \ 

\newpage

\onecolumngrid
\begin{center}
    {\bf \large End Matter}
\end{center}

\twocolumngrid
The adsorption energy of a TM adatom on a FE substrate is defined as $E_{\mathrm{ads}} = E(\mathrm{HS}) - E(\mathrm{FE}) - E(\mathrm{TM})$.
A more negative $E_{\mathrm{ads}}$ indicates stronger binding.
For a Cr adatom on \(\mathrm{In_2Se_3}\), we consider three adsorption geometries: Hollow, Top--In, and Top--Se \cite{nc1}.
Notably, for both polarization states of \(\mathrm{In_2Se_3}\), the Hollow site yields the lowest $E_{\mathrm{ads}}$ (see Table II), i.e., it is the most stable configuration. 
The energy mapping methods \cite{ssm12, sssm12} are then applied to resolve the magnetic parameters within a single dimer [see crystal structures in Fig.~\ref{fig:P-up-dn}~(a)]. 
Specifically, the Heisenberg exchange is obtained via $J=(E_{\rm AFM}-E_{\rm FM})/2S^{2}$; magnetic anisotropy is determined from energy differences between different magnetization orientations: $K_1 = (E_x - E_z)/2 S^2$ and $K_2 = (E_y - E_z)/2 S^2$; and the DMI is obtained from the energy difference between opposite spin chiralities: $D = (E_{acw} - E_{cw})/2 S^2$.
In DMI calculations, the spin canting angle between nearest neighbors is fixed at 90°, and the DMI components along different directions are evaluated by rotating the spins in planes perpendicular to the corresponding DMI directions.
Relativistic effects are included in all magnetic parameter calculations. 
\begin{table}[h!]
\centering
\setlength{\tabcolsep}{11pt} 
\renewcommand{\arraystretch}{1}
\begin{tabular}{cccccc}
\toprule
\multicolumn{2}{c}{Hollow site} & \multicolumn{2}{c}{Top--In site} & \multicolumn{2}{c}{Top--Se site} \\
\cmidrule(lr){1-2}\cmidrule(lr){3-4}\cmidrule(lr){5-6}
P$_\text{dn}$ & P$_\text{up}$ & P$_\text{dn}$ & P$_\text{up}$ & P$_\text{dn}$ & P$_\text{up}$ \\
\midrule
-3.51 & -2.32 & -3.45 & -2.18 & -2.57 & -1.79 \\
\bottomrule
\end{tabular}
\caption{The binding energy of the Cr atom in the FE-magnetic HS Cr/In$_2$Se$_3$ system. All values are in eV.} 
\label{tab:parameters}
\end{table}

In 2D magnets, the ligand-mediated superexchange could render next-nearest- and further-neighbor couplings non-negligible \cite{a1}.
Therefore, to ensure the Cr dimer remains isolated, the dimer is placed in a large 5×5 In$_2$Se$_3$ supercell in all calculations.
The center-to-center distance between dimers is 21 Å, much larger than the 4 Å Cr–Cr separation within the dimer.
We then double the supercell along the $\bf{a}$ axis (2×1) to include two Cr dimers.
By comparing FM and AFM alignments of adjacent Cr dimers in this 2×1 supercell, we deduce an inter-dimer exchange coupling $J =(E_{\rm AFM} - E_{\rm FM})/8 S^2$ of -0.005 to -0.011 meV for P$_\text{dn}$ and P$_\text{up}$ respectively, which is several orders of magnitude smaller than the intradimer exchange [Table I of main text]. 
 

\begin{thebibliography}{99}

\bibitem{Eriksson2017} O. Eriksson, A. Bergman, L. Bergqvist, and J. Hellsvik,
Atomistic Spin Dynamics: Foundations and Applications
(Oxford University Press, Oxford, 2017).

\bibitem{Lakshmanan2011} M. Lakshmanan, 
The fascinating world of the Landau–Lifshitz–Gilbert equation: an overview,
Philos. Trans. R. Soc. A {\bf 369}, 1280 (2011).

\bibitem{gaitan24} F. Garcia-Gaitan and B. K. Nikoli\'c, 
Fate of entanglement in magnetism under Lindbladian or non-Markovian dynamics and conditions for their transition to Landau-Lifshitz-Gilbert classical dynamics, 
Phys. Rev. B {\bf 109}, L180408 (2024). 

\bibitem{uhrig25} G. S. Uhrig, 
Landau-Lifshitz damping from Lindbladian dissipation in quantum magnets, 
New J. Phys. {\bf 27}, 103502 (2025). 

\bibitem{verstraten23} R. C. Verstraten, T. Ludwig, R. A. Duine, and C. Morais Smith, 
Fractional Landau-Lifshitz-Gilbert equation, 
Phys. Rev. Research {\bf 5}, 033128 (2023). 

\bibitem{Wieser2013} R. Wieser,
Comparison of quantum and classical relaxation in spin dynamics,
Phys. Rev. Lett. {\bf 110}, 147201 (2013).

\bibitem{liu24} Y. Liu, I. P. Miranda, L. Johnson, A. Bergman, A. Delin, D. Thonig, M. Pereiro, O. Eriksson, V. A. Mousolou, and E. Sj\"oqvist,
Quantum Analog of Landau-Lifshitz-Gilbert Dynamics,
Phys. Rev. Lett. {\bf 133}, 266704 (2024).

\bibitem{yuan22} H. Y. Yuan, Y. Cao, A. Kamra, R. A. Duine, and P. Yan,  
Quantum magnonics: When magnon spintronics meets quantum information science, 
Phys. Rep. {\bf 965}, 1 (2022). 

\bibitem{kamra20} A. Kamra, W. Belzig, and A. Brataas,
Magnon-squeezing as a niche of quantum magnonics,
Appl. Phys. Lett. {\bf 117}, 090501 (2020).

\bibitem{romling23} A.-L. E. R\"omling, A. Vivas-Via\~na, C. S\'anchez Mu\~noz, and A. Kamra,
Resolving nonclassical magnon composition of a magnetic ground state via a qubit,
Phys. Rev. Lett. {\bf 131}, 143602 (2023). 

\bibitem{romling24} A.-L. E. R\"omling and A. Kamra,
Quantum sensing of antiferromagnetic magnon two-mode squeezed vacuum,
Phys. Rev. B {\bf 109}, 174410 (2024). 

\bibitem{yuan22L} H. Y. Yuan, W. P. Sterk, A. Kamra, and R. A. Duine,
Pure dephasing of magnonic quantum states,
Phys. Rev. B {\bf 106}, L100403 (2022). 

\bibitem{yuan22ME} H. Y. Yuan, W. P. Sterk, A. Kamra, and R. A. Duine,
Master equation approach to magnon relaxation and dephasing,
Phys. Rev. B {\bf 106}, 224422 (2022). 

\bibitem{scheie21} A. Scheie, P. Laurell, A. M. Samarakoon, B. Lake, S. E. Nagler, G. E. Granroth, S. Okamoto, G. Alvarez, and D. A. Tennant,
Witnessing entanglement in quantum magnets using neutron scattering,
Phys. Rev. B {\bf 103}, 224434 (2021); see also Phys. Rev. B {\bf 107}, 059902(E) (2023).

\bibitem{me1} D. Pantel, S. Goetze, D. Hesse, and M. Alexe, Reversible electrical switching of spin polarization in multiferroic tunnel junctions, 
Nat. Mater. {\bf 11}, 289 (2012).

\bibitem{gong1} C. Gong, E. M. Kim, Y. Wang, G. Lee, and X. Zhang, Multiferroicity in atomic van der Waals heterostructures, 
Nat. Commun. {\bf 10}, 2657 (2019).

\bibitem{gong2} S. Liang, T. Xie, N. A. Blumenschein, T. Zhou, T. Ersevim, Z. Song, J. Liang, M. A. Susner, B. S. Conner, S.-J. Gong, J.-P. Wang, M. Ouyang, I. Zutic, A. L. Friedman, X. Zhang, and C. Gong, Small-voltage multiferroic control of two-dimensional magnetic insulators, 
Nat. Electron. {\bf 6}, 199 (2023). 

\bibitem{fe2} J. H. Lee, H. J. Kim, J. Yoon, S. Kim, J. R. Kim, W. Peng, S. Y. Park, T. W. Noh, and D. Lee, 
Flexoelectricity-Driven Mechanical Switching of Polarization in Metastable Ferroelectrics, 
Phys. Rev. Lett. {\bf 129}, 117601 (2022).

\bibitem{fe3} Q. Yang and S. Meng, Light-Induced Complete Reversal of Ferroelectric Polarization in Sliding Ferroelectrics, 
Phys. Rev. Lett. {\bf 133}, 136902 (2024).

\bibitem{Fert2024} A. Fert, R. Ramesh, V. Garcia, F. Casanova, and M. Bibes,
Electrical control of magnetism by electric field and current-induced torques,
Rev. Mod. Phys. {\bf 96}, 015005 (2024).

\bibitem{stm1} D. Eigler and E. Schweizer, Positioning single atoms with a scanning tunnelling microscope,  
Nature (London) {\bf 344}, 524 (1990). 

\bibitem{stm2} C. F. Hirjibehedin, C. P. Lutz, and A. J. Heinrich, 
Spin Coupling in Engineered Atomic Structures, 
Science {\bf 312}, 1021 (2006).

\bibitem{stm3} S. Loth, S. Baumann, C. P. Lutz, D. M. Eigler, and A. J. Heinrich, Bistability in Atomic-Scale Antiferromagnets, Science {\bf 335}, 196 (2012).

\bibitem{stm4} S. Kawai, A. S. Foster, F. F. Canova, H. Onodera, S. Kitamura, and E. Meyer, 
Atom manipulation on an insulating surface at room temperature, 
Nat. Commun. {\bf 5}, 4403 (2014). 

\bibitem{sm2} G. Kresse and J. Hafner,
Ab initio molecular-dynamics simulation of the liquid-metal–amorphous-semiconductor transition in germanium,
Phys. Rev. B {\bf 49}, 14251 (1994).

\bibitem{sm3} G. Kresse and J. Furthmüller, 
Efficient iterative schemes for ab initio total-energy calculations using a plane-wave basis set, 
Phys. Rev. B {\bf 54}, 11169 (1996).

\bibitem{sm4} G. Kresse and D. Joubert, 
From ultrasoft pseudopotentials to the projector augmented-wave method, 
Phys. Rev. B {\bf 59}, 1758 (1999).

\bibitem{sm5} J. P. Perdew, K. Burke, and M. Ernzerhof, 
Generalized Gradient Approximation Made Simple, 
Phys. Rev. Lett. {\bf 77}, 3865 (1996).

\bibitem{sm6} W. Ding, J. Zhu, Z. Wang, Y. Gao, D. Xiao, Y. Gu, Z. Zhang, and W. Zhu, 
Prediction of intrinsic two-dimensional ferroelectrics in In$_2$Se$_3$ and other III$_2$-VI$_3$ van der Waals materials, 
Nat. Commun. {\bf 8}, 14956 (2017). 

\bibitem{sm7} Y. Zhou, D. Wu, Y. Zhu, Y. Cho, Q. He, X. Yang, K. Herrera, Z. Chu, Y. Han, M. C. Downer, H. Peng, and K. Lai, 
Out-of-Plane Piezoelectricity and Ferroelectricity in Layered $\alpha$-In$_2$Se$_3$ Nanoflakes, 
Nano Lett. {\bf 17}, 5508 (2017). 

\bibitem{sm8} J. Xiao, H. Zhu, Y. Wang, W. Feng, Y. Hu, A. Dasgupta, Y. Han, Y. Wang, D. A. Muller, L. W. Martin, P. Hu, and X. Zhang,  
Intrinsic Two-Dimensional Ferroelectricity with Dipole Locking, 
Phys. Rev. Lett. {\bf 120}, 227601 (2018). 

\bibitem{a1} D. Amoroso, P. Barone, and S. Picozzi, 
Spontaneous skyrmionic lattice from anisotropic symmetric exchange in a Ni-halide monolayer, 
Nat. Commun. {\bf 11}, 5784 (2020). 

\bibitem{sm9} B. Goodenough,
Theory of the role of covalence in the perovskite-type manganites [La, M(II)]MnO3,
Phys. Rev. {\bf 100}, 564 (1955). 

\bibitem{sm10} J. Kanamori,
Superexchange interaction and symmetry properties of electron orbitals,
J. Phys. Chem. Solids {\bf 10}, 87 (1959).  

\bibitem{sm11} P. W. Anderson, 
New approach to the theory of superexchange interactions, 
Phys. Rev. {\bf 115}, 2 (1959).

\bibitem{ssm11} Q. Cui, J. Liang, Z. Shao, P. Cui, H. Yang, 
Strain-tunable ferromagnetism and chiral spin textures in two-dimensional Janus chromium dichalcogenides, 
Phys. Rev. B {\bf 102}, 094425 (2020).  

\bibitem{ssm12} H. Yang, J. Liang, Q. Cui, 
First-principles calculations for Dzyaloshinskii–Moriya interaction, 
Nat. Rev. Phys. {\bf 5}, 43 (2023).

\bibitem{sssm12} A. Edstr\"om, D. Amoroso, S. Picozzi, P. Barone , and M. Stengel, 
Curved Magnetism in CrI$_3$, 
Phys. Rev. Lett. {\bf 128}, 177202 (2022).

\bibitem{sm12} Q. Cui, Y. Zhu, J. Jiang, J. Liang, D. Yu, P. Cui, and H. Yang, 
Ferroelectrically controlled topological magnetic phase in a Janus-magnet-based multiferroic heterostructure, 
Phys. Rev. Research {\bf 3}, 043011 (2021).

\bibitem[]{remark-vNEntr}
The von Neumann entropy is defined as
\begin{equation*}
S_{\rm vN}(\varrho_A) = -\mathrm{Tr} ( \varrho_A \ln \varrho_A ) = -\sum_i \lambda_i \ln \lambda_i \, ,
\end{equation*}
where $\varrho_A = \mathrm{Tr}_B\bigl(|\psi\rangle\langle\psi|\bigr)$ is the reduced density matrix of subsystem $A$ obtained by tracing out its complement $B$, and $\{\lambda_i\}$ are the eigenvalues of $\varrho_A$.
Because $\varrho_A$ is positive semi-definite with unit trace, $S(\varrho_A)=0$ if and only if $|\psi\rangle$ is a product state, while $S(\varrho_A)$ attains its maximum value $\log_2 d$ when $|\psi\rangle$ is maximally entangled in dimension $d$. Consequently, under the purity-preserving \cite{liu24} $q$-LLG evolution, the von Neumann entropy serves as a faithful and easily computable entanglement monotone.
For a spin-2 dimer, for which $d=2s+1=5$, the von Neumann entropy cannot exceed $\log_2 5 \approx 2.32$, the value attained by a maximally entangled state.

\bibitem{nc1} L. Ju, X. Tan, X. Mao, Y. Gu, S. Smith, A. Du, Z. Chen, C. Chen, and L. Kou, 
Controllable CO$_2$ electrocatalytic reduction via ferroelectric switching on single atom anchored In$_2$Se$_3$ monolayer, 
Nat. Commun. {\bf 12}, 5128 (2021).

\bibitem{prls1}  D. Yu, Y. Ga, J. Liang, C. Jia , and H. Yang, 
Voltage-Controlled Dzyaloshinskii-Moriya Interaction Torque Switching of Perpendicular Magnetization, Phys. Rev. Lett. {\bf 130}, 056701 (2023).

\bibitem{nl1} Z. Xia, O. V. Prezhdo, W. Zhu, J. Zhao, and Q. Zheng, 
Picosecond Valley Manipulation in Two-Dimensional In2Se3 via Ferroelectric Switching, 
Nano Lett. {\bf 25}, 9084 (2025).

\bibitem{Hartmann2025NonMarkov} F. Hartmann, V. Unikandanunni, M. Bargheer, E. E. Fullerton, S. Bonetti, and J. Anders,
Intrinsic non-Markovian magnetisation dynamics,
arXiv:2512.07378 (2025).

\bibitem{Felipe2025} F. Reyes-Osorio and B.  K. Nikolić,
Optically induced magnetic inertia and magnons from non-markovian extension of the Landau-Lifshitz-Gilbert Equation,
Phys. Rev. Lett. {\bf 135}, 246701 (2025).

\bibitem{ke1} A. Kirilyuk, A. V. Kimel, and T. Rasing, 
Ultrafast optical manipulation of magnetic order, 
Rev. Mod. Phys. {\bf 82}, 2731 (2010).

\bibitem{ke2} M. J. Gomez, K. Liu, J. G. Lee, and R. B. Wilson, 
High sensitivity pump–probe measurements of magnetic, thermal, and acoustic phenomena with a spectrally tunable oscillator, 
Rev. Sci. Instrum. {\bf 91}, 023905 (2020).

\end{thebibliography}
\end{document}